\documentclass[prl,floatfix,twocolumn,showpacs,amsmath,amssymb,superscriptaddress]{revtex4}

\usepackage{graphicx}

\usepackage{bm}
\usepackage{latexsym}
\usepackage{amssymb}
\usepackage{amsfonts}
\usepackage{amsmath}
\usepackage{times}
\usepackage[colorlinks,bookmarks=false,citecolor=blue,linkcolor=red,urlcolor=blue]{hyperref}

\begin{document}

\author{Tam\'as A. T\'oth}
\affiliation{Institut de th\'eorie des ph\'enom\`enes physiques,
Ecole Polytechnique F\'ed\'erale de Lausanne, CH-1015 Lausanne, Switzerland}
\author{Andreas M. L\"auchli}
\affiliation{Max Planck Institut f\"ur Physik komplexer Systeme, D-01187 Dresden, Germany}
\author{Fr\'ed\'eric Mila}
\affiliation{Institut de th\'eorie des ph\'enom\`enes physiques,
Ecole Polytechnique F\'ed\'erale de Lausanne, CH-1015 Lausanne, Switzerland}
\author{Karlo Penc}
\affiliation{Research Institute for Solid State Physics and Optics, H-1525 Budapest, P.O. Box 49, Hungary}

\date{\today}

\title{Three-sublattice ordering of the SU(3) Heisenberg model of three-flavor fermions\\
 on the square and cubic lattices}

\begin{abstract}
Combining a semi-classical analysis with exact diagonalizations, we
show that the ground state of the SU(3) Heisenberg model on the square lattice develops three-sublattice long-range order.
This surprising pattern for a bipartite lattice with only nearest-neighbor interactions is shown to be the consequence of a subtle quantum order-by-disorder mechanism.
By contrast, thermal fluctuations favor two-sublattice configurations via entropic selection. These results are shown to extend to the cubic lattice, and experimental implications
for the Mott-insulating states of three-flavor fermionic atoms in optical lattices are discussed.
\end{abstract}


\maketitle

Mott transitions, i.e. metal--insulator transitions driven by correlations, and the nature of the associated
Mott insulating phases represent one of the central themes of contemporary condensed matter physics~\cite{mitImada},
and more recently also of the field of ultracold atomic gases~\cite{blochRMP}.
Theoretically, the canonical case of two-flavor fermions on hypercubic lattices is thoroughly understood. For strong interactions 
an antiferromagnetically ordered two--sublattice N\'eel state is realized. Ongoing experimental efforts using
ultracold fermionic gases are focused on reaching this state coming from higher temperatures~\cite{twoflavorMott}.

In an exciting parallel development, recent experimental advances using multi-flavor 
atomic gases~\cite{threeflavorLi,SrBEC} have paved the way to the investigation of Mott insulating states with more than two flavors in optical lattices~\cite{honerkamp2004,groshkov2010}.
While it is intuitively clear that Mott insulating states will exist at particular commensurate fillings -- as suggested by atomic limit considerations and
single site DMFT simulations~\cite{gorelik2009,miyatake2010} --,
the nature and the spatial structure of multi-flavor Mott insulating states are in general not well understood. For instance, on the square lattice geometry,
many different proposals for insulating states have been put forward, ranging from SU($N$) symmetry breaking "magnetic" states to dimerized or plaquette states,
chiral spin liquids and staggered flux phases~\cite{affleck1988,marston1989,vandenbossche2000,honerkamp2004,FaWang2009}.

In this Letter, we present a strong case in favor of a three-sublattice long-range ordered ground state
for the Mott insulating state of three-flavor $(N=3)$ fermions with one particle per site (1/3--filling)
on the square lattice. This is based on analytical and numerical investigations of the strong coupling limit $U\gg t$ of the SU$(3)$ symmetric Hubbard model defined by the Hamiltonian
\begin{equation}
\mathcal{H} = -t \sum_{\langle i,j\rangle,\alpha} (c^\dagger_{i,\alpha} c^{\phantom{\dagger}}_{j,\alpha} + \mbox{h.c.})
+ U \sum_{i,\alpha<\beta} n_{i,\alpha} n_{i,\beta} \;.
\end{equation}
Here $c^\dagger_{i,\alpha}$ and  $c^{\phantom{\dagger}}_{i,\alpha}$ create and annihilate a fermion at site $i$ with flavor $\alpha$, respectively, and $n_{i,\alpha} = c^\dagger_{i,\alpha} c^{\phantom{\dagger}}_{i,\alpha}$.
To second order in $t/U$, the low-energy physics is captured by the SU(3) antiferromagnetic Heisenberg model with
coupling $J=2t^2/U$:
\begin{eqnarray}
  \mathcal{H} =  J \sum_{\langle i,j\rangle} \mathcal{P}_{ij} \;,
  \label{eq:HSU3}
\end{eqnarray}
where $\mathcal{P}_{ij}$ is a transposition operator that exchanges SU(3) spins on site $i$ and $j$:
$\mathcal{P}_{ij} |\alpha_i \beta_j \rangle = |\beta_i\alpha_j\rangle$. The spins on a site form
the 3-dimensional fundamental irreducible representation of the SU(3) algebra. In the following, the
basis states will be denoted by $\vert A\rangle$, $\vert B\rangle$ and $\vert C\rangle$. Note that this
model can also be seen as a special high--symmetry point of the SU(2) spin--1 bilinear--biquadratic exchange
Hamiltonian when bilinear and biquadratic couplings are equal.

\begin{figure}[t]
\includegraphics[width=6 truecm ]{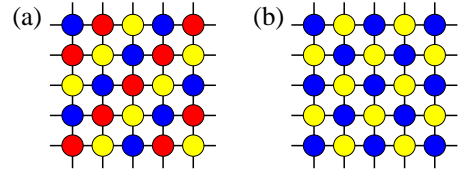}
\caption{(Color online) Sketch of the (a) three-sublattice and (b) two-sublattice phases of the SU(3) antiferromagnetic Heisenberg model.
\label{fig:phases}}
\end{figure}

Let us start by briefly reviewing what is known about this model in various geometries. In one dimension, the model
has a Bethe Ansatz solution~\cite{BetheAnsatz}. It has gapless excitations at $q=0$ and $q=\pm 2\pi/3$ \cite{sutherland}, and the correlations decay algebraically with period 3. In higher dimensions, much less is known, and most of it relies on the
pioneering work of Papanicolaou~\cite{papanicolaou88} who has investigated this question in the context
of spin--1 models with the help of a variational approach, using a site--factorized wavefunction of the form
\begin{equation}
 | \Psi \rangle =   \prod_{i=1}^{N_\Lambda} \left(d_{A,i} |A\rangle_i + d_{B,i} |B\rangle_i + d_{C,i} |C\rangle_i \right) \;,
  \label{eq:Psivari}
\end{equation}
where $N_\Lambda$ is the number of sites.
Grouping the variational parameters into (complex) vectors $\mathbf{d}_i = (d_{A,i}, d_{B,i}, d_{C,i})$ and imposing the normalization $\mathbf{d}_i \cdot \mathbf{\bar d}_i = 1$, the problem reduces to the minimization of
\begin{equation}
E_{\rm var}=
\frac{\langle\Psi|\mathcal{H}|\Psi \rangle}{\langle\Psi |\Psi \rangle} =  J \sum_{\langle i,j\rangle} \left|\mathbf{d}_i \cdot \mathbf{\bar d}_j \right|^2
\;.
\end{equation}
Since $J>0$, the energy of nearest-neighbor bonds is minimal when the $\mathbf{d}$ vectors, hence the wavefunctions,
are orthogonal. The nature of the variational ground state then depends on the connectivity of the lattice. For
a triangle, this condition enforces three mutually orthogonal $\mathbf{d}$ vectors. As a consequence,
for the triangular lattice, the energy is minimized up to global SU(3) rotations by a single wavefunction
constructed by choosing on each of the three sublattices the wavefunctions $|A\rangle$, $|B\rangle$, and $|C\rangle$,
respectively. The three-sublattice long-range order embodied by this wavefunction has recently been shown
to be stable against quantum fluctuations~\cite{triangularT}.

By contrast, the square lattice does not provide enough constraints to uniquely select a set of mutually
orthogonal $\mathbf{d}$ vectors. Consider for instance a N\'eel state with $|A\rangle$ and $|B\rangle$
on the two sublattices. Any state obtained by replacing $|B\rangle$ by $|C\rangle$ on an arbitrary number
of sites is also a ground state. This variational approach thus leads to a highly degenerate ground-state
manifold. This situation is reminiscent of frustrated SU(2) antiferromagnetism, where the competition
between exchange paths often leads to an infinite number of classical ground states: in that case, quantum
or thermal fluctuations often restore long-range order by a selection mechanism that favors collinear
or planar configurations and is known as 'order-by-disorder'~\cite{henley1989}. For the SU(3) model, zero-point quantum fluctuations can be calculated with the help of the flavor-wave theory,
an extension of the SU(2) spin-wave theory to the SU(3) case \cite{papanicolaou88, joshi}.
This approach starts from the representation of the model in terms of three-flavor Schwinger bosons:
\begin{equation}
\mathcal{P}_{ij} = \sum_{\mu,\nu\in\{A,B,C\}} a^{\dagger}_{\mu,i} a^{\dagger}_{\nu,j}
a^{\phantom{\dagger}}_{\nu,i}  a^{\phantom{\dagger}}_{\mu,j} \;,
\label{eq:PwSB}
\end{equation}
with the constraint $\sum_\nu a^\dagger_{\nu,i} a^{\phantom{\dagger}}_{\nu,i}=1$.
In order to treat quantum fluctuations around a variational solution defined by $\mathbf{d}_j$, one first performs a local SU(3) rotation of the Schwinger bosons by choosing two vectors $\mathbf{e}_j$ and $\mathbf{f}_j$ which, together with $\mathbf{d}_j$, define a local orthogonal basis, in terms of which the rotated bosons are given as
$\tilde a^{\dagger}_{A,j} = \sum_\mu d_{\mu,j} a^{\dagger}_{\mu,j}$,
$\tilde a^{\dagger}_{B,j} = \sum_\mu e_{\mu,j} a^{\dagger}_{\mu,j}$, and
$\tilde a^{\dagger}_{C,j} = \sum_\mu f_{\mu,j} a^{\dagger}_{\mu,j}$.
A 'semiclassical' $1/M$  expansion is then generated by replacing the constraint by $\sum_\nu a^\dagger_{\nu,i} a^{\phantom{\dagger}}_{\nu,i}=M$ and the Schwinger bosons along the local direction of the variational solution by
\begin{equation}
\tilde a^{\dagger}_{A,i},\tilde a^{\phantom{\dagger}}_{A,i} \rightarrow
\sqrt{
M
- \tilde a^{\dagger}_{B,i} \tilde a^{\phantom{\dagger}}_{B,i}
- \tilde a^{\dagger}_{C,i} \tilde a^{\phantom{\dagger}}_{C,i}} \;.
 \label{eq:ayfw}
\end{equation}
The bosons $\tilde a_{B,i}$ and $\tilde a_{C,i}$ play the role of Holstein-Primakoff bosons.
Expanding in powers of $1/M$ leads to
\begin{equation}
\mathcal{P}_{ij}=
M\left(
\mathbf{d}_j \cdot \mathbb{A}^{\dagger}_i
+ \mathbf{\bar d}_i \cdot \mathbb{A}^{\phantom{\dagger}}_j
\right)
\left(
\mathbf{\bar d}_j \cdot \mathbb{A}^{\phantom{\dagger}}_i
+ \mathbf{d}_i \cdot \mathbb{A}^{\dagger}_j
\right)-M
\label{eq:PdefwHP}
\end{equation}
with
$\mathbb{A}^{\phantom{\dagger}}_i=\mathbf{e}_i \tilde a^{\phantom{\dagger}}_{B,i} + \mathbf{f}_i \tilde a^{\phantom{\dagger}}_{C,i}$. The resulting Hamiltonian is quadratic and can be diagonalized by a Bogoliubov
transformation. The zero-point energy is half the sum of the eigenfrequencies.

\begin{figure*}[t]
\includegraphics[width= 16truecm]{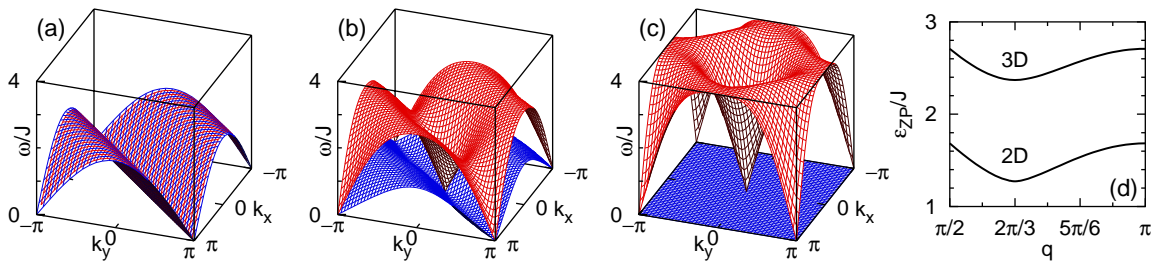}
\caption{(Color online) Flavor--wave dispersions of (a) the three-sublattice state, (b) the helical state with $\cos \theta = 0.4$, and (c) the two-sublattice state.
(d) Zero-point energy per site, $\varepsilon_\text{ZP}=(1/N_\Lambda) \sum_{\mathbf{k},\nu} \omega_\nu(\mathbf{k})/2$, as a function of the wavevector $(q,\pm q)$ of the helix.
The minimum is located at $q=2\pi/3$ in both two and three dimensions.  \label{fig:disps}}
\end{figure*}

Now, simple considerations lead to two natural candidates for possible orderings: (i) the two--sublattice state with ordering wavevector $\mathbf{Q}_2=(\pi,\pi)$, as suggested by the bipartite nature of the lattice; (ii) a three-sublattice state with ordering wavevector $\mathbf{Q}_3^\pm=(2\pi/3,\pm 2\pi/3)$, as suggested by the softening of the excitation spectrum in one--dimensional chains. Both states feature diagonal stripes
of equal spins, with an alternation of A and B stripes in case (i) and a succession of A, B and C stripes in case (ii) (see Fig.~\ref{fig:phases}). For such states, the excitation spectrum has the periodicity of the square lattice,
leading to
\begin{equation}
\mathcal{H} = - 2 M J N_\Lambda + M \sum_{\nu=1}^2 \sum_\mathbf{k} \omega_{\nu}(\mathbf{k})
\left(\alpha_\nu^{\dagger} (\mathbf{k}) \alpha_\nu (\mathbf{k}) +\frac{1}{2}\right) \;.
\end{equation}
The spectrum consists of two branches because there are two states orthogonal to the local variational
state. For the three-sublattice state, the branches are degenerate with dispersion
\begin{equation}
\omega_{1,2}(\mathbf{k}) = 2 J \sqrt{1-|\gamma_\mathbf{k}|^2} \;,
\label{eq:omega3}
\end{equation}
while for the two-sublattice case they are given by
\begin{equation}
\omega_1(\mathbf{k}) = 2 J \sqrt{4-\left(\gamma_\mathbf{k}+\bar\gamma_\mathbf{k}\right)^2}\;, \quad \omega_2(\mathbf{k})=0\;,
\label{eq:omega2}
\end{equation}
with $\gamma_\mathbf{k} = (e^{ik_x}+e^{ik_y})/2$.
The corresponding spectra are shown in Fig.~\ref{fig:disps}(a) and (c), respectively.

In both cases, there is an infinite number of zero modes. For the two-sublattice case, the presence
of an entirely soft branch is natural: if the two-sublattice variational state
corresponds, say, to A and B spins, each spin can tilt toward C at no energy cost. What is more surprising is the
line of zero modes in the three-sublattice case, since one would naively expect zero modes only at $\mathbf{k}=0,\pm\mathbf{Q}_3^{\pm}$. This suggests that there must be a family of classical ground states with wave vectors $(q,\pm q)$.
Indeed, after choosing
the vectors $\mathbf{d}_{l}$ and $\mathbf{d}_{l+1}$ of two consecutive stripes, we may write the next one as $\mathbf{d}_{l+2} = \cos\theta \mathbf{d}_{l} + \sin\theta \mathbf{d}_{l}\times\mathbf{\bar d}_{l+1}$. If we continue
this construction using always the same $\theta$, we end up with a helical state of the form
$|\psi_i \rangle \propto
\sqrt{-\cos q} |u \rangle + \sin q(x_i+y_i) |v\rangle + \cos q(x_i+y_i) |w\rangle $
with  $\cos\theta = -1 - 2\cos q$ and $\pi/2<q<\pi$, where $\vert u\rangle,\vert v\rangle,\vert w\rangle$ depend on the initial choice of $\mathbf{d}_{l}$ and $\mathbf{d}_{l+1}$. These helical states extrapolate between
the two-sublattice case ($q=\pi/2$ and $q=\pi$) and the three-sublattice case ($q=2\pi/3$). Their excitations have the periodicity of the lattice, and the dispersion is given
by the non-negative solutions of the equation (extended to a cubic lattice of arbitrary dimension $D$ for later
reference)
\begin{eqnarray}
&&\omega^4 -D^2 J^2 \left[2 (1-\gamma_\mathbf{k} \bar\gamma_\mathbf{k})+(2-\gamma_\mathbf{k}^2-\bar\gamma_\mathbf{k}^2)\cos^2\theta\right] \omega^2 +\nonumber\\
&& D^4 J^4 \sin^4 \theta (1-\gamma_\mathbf{k}\bar\gamma_\mathbf{k})^2= 0 \;,
\label{quantum_modes}
\end{eqnarray}
where $\gamma_\mathbf{k} = (e^{ik_x}+e^{ik_y}+\dots)/D$. An example is shown in Fig.~\ref{fig:disps}(b). All helical
states give rise to a line of zero modes.

We see therefore that there is in fact an infinite number of helical ground states that are {\it a priori} good candidates for being the quantum ground state. We have calculated and compared the zero-point energy of these states (Fig.~\ref{fig:disps}), and found that it is minimal for $q=2\pi/3$. 
We have also compared the zero-point energy of the
three-sublattice state with that of random ground states on finite clusters,
as well as with that of stripe states with lower periodicity, with the conclusion that it is always lower. So, within the flavor-wave theory, the stabilization of the three-sublattice state appears to be quite robust. This leads to the first important conclusion
of this paper: on the square lattice, quantum fluctuations stabilize three-sublattice long-range order. 

This conclusion is quite surprising from the point of view of order-by-disorder. Indeed, the two-sublattice state has by far the largest number of zero modes, and according to common wisdom it should be selected. However, this need not be the case for quantum fluctuations: if the non-zero modes have sufficiently large energy, they may compensate for the vanishing
contribution of the zero modes. This is what happens here for the two-sublattice structure, whose upper
branch is larger than twice the degenerate three-sublattice branch for all wavevectors.
By contrast, for thermal fluctuations, the stabilization of the configuration with the largest number of zero
modes is systematic, since the low-temperature free energy reads \cite{chalker1992}
\begin{equation}
F=E_0-\frac{N_\text{ZM}}{4}T\ln T-\frac{N_M-N_\text{ZM}}{2}T\ln T \;,
\end{equation}
where $N_M$ is the total number of modes and $N_{ZM}$ is the number of zero modes.
The classical spectrum $\lambda$ is found by diagonalizing the quadratic form obtained
by replacing creation and annihilation operators by complex numbers. For the helical states, its four branches 
are given by 
\begin{equation}
  \lambda^2 - D J \left[2\pm \cos\theta (\gamma_{\mathbf{k}}+ \bar\gamma_{\mathbf{k}})\right] \lambda
  + D^2 J^2 \sin^2\theta \left(1-\gamma_{\mathbf{k}} \bar\gamma_{\mathbf{k}}\right) = 0 \;.
  \label{thermal_modes}
\end{equation}
Similarly to the SU(2) case, the classical and quantum spectra are different
but related, and in dimension $D\ge 2$, the proportion of zero modes is the same. This analysis predicts therefore that thermal fluctuations stabilize the two-sublattice state, in agreement with
Classical Monte Carlo simulations ~\cite{BCL_unpublished}.

\begin{figure}[t]
\includegraphics[width= 7.5truecm]{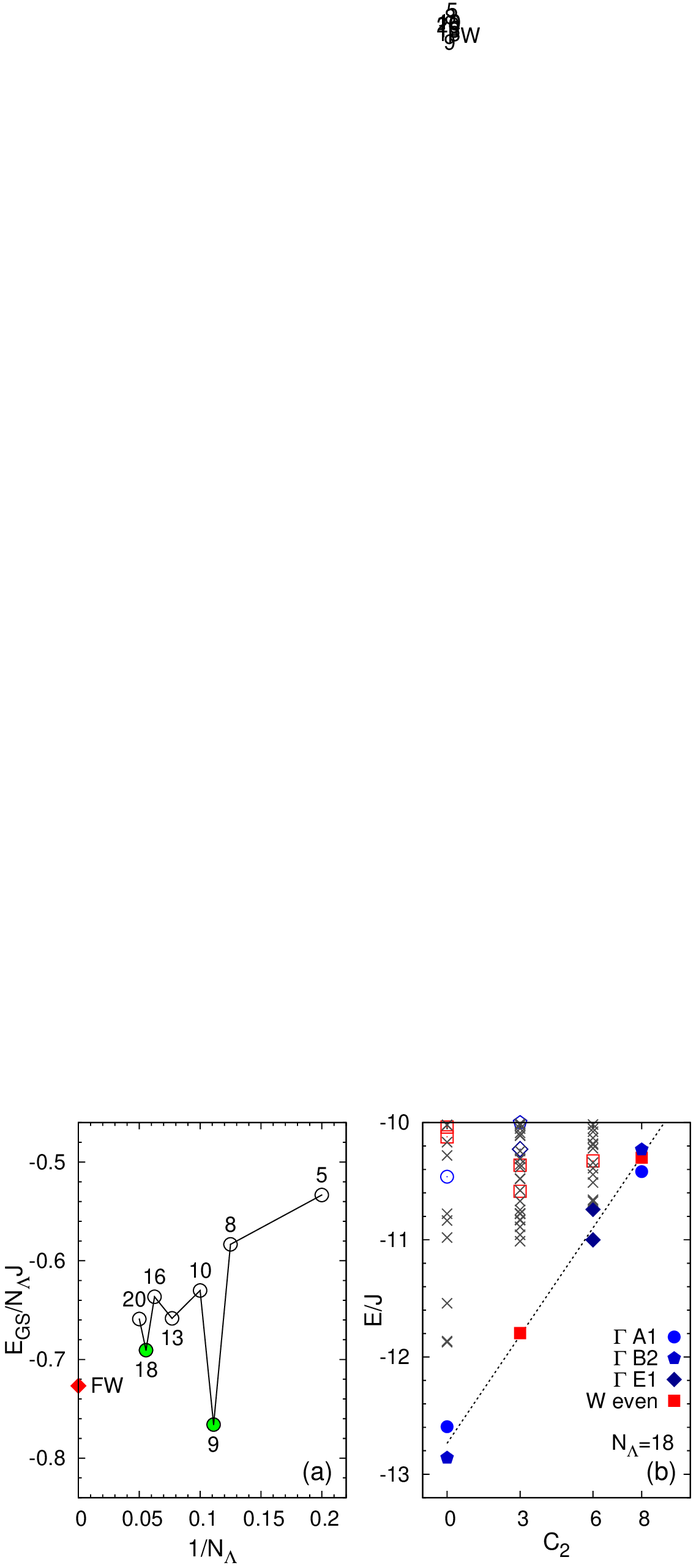}
\caption{(Color online)
(a) Energy per site from exact diagonalization of various finite size square clusters compared to the flavor-wave (FW) result. (b) Tower of states for 18 sites. $\Gamma$ denotes points in the center of the Brillouin zone, with A1 and A2 one--dimensional and E1 two--dimensional irreps, while W is a four dimensional irrep with wave vectors $(\pm 2\pi/3, \pm 2\pi/3)$. The tower
contains the irreducible representations expected for a striped three-sublattice flavor-wave state with two possible
orientations.
 \label{fig:energypersite_tower}}
\end{figure}

Next, we compare these predictions with exact diagonalizations (ED) of the Hamiltonian of Eq.~(\ref{eq:HSU3})
on finite clusters. The energy per
site for square samples of up to 20 sites is shown in Fig.~\ref{fig:energypersite_tower}(a).
It is significantly smaller for the samples whose number of sites is a multiple of 3 (9 and 18),
providing evidence in favor of a three-sublattice symmetry breaking. To check if the continuous
SU(3) symmetry is also broken, we have plotted in Fig.~\ref{fig:energypersite_tower}(b) the energy levels as a function
of the quadratic Casimir operator $C_2$ of SU(3), keeping track of the irreducible representations of the
space group symmetry. In the case of the continuous symmetry breaking one expects the low-energy part of the
spectrum to align linearly as a function of $C_2$, giving rise to a {\em tower of states}~\cite{PencSU4triangular,PencLauchliChapter}.
This is clearly the case in Fig.~\ref{fig:energypersite_tower}(b), as highlighted by the dashed line,  
This tower of states can be thought of as a combination of two towers corresponding to the two
possible propagation vectors ($\mathbf{Q}_3^\pm$), which results into the finite-size splitting of
some levels (e.g. $\Gamma A1$ and $\Gamma B2$), as well as the increased degeneracy of some irreps (e.g. $W$).
Note also that the tower is not as well separated from other states as in other systems~\cite{PencSU4triangular},
a consequence of the order-by-disorder selection mechanism that leads to low-lying excitations
associated to other mean-field solutions.
The structure of the energy spectrum further indicates that the state with an equal population of the SU(3) basis states is
stable with respect to the occurrence of spontaneous population imbalance or phase separation.
Finally, an inspection of the real-space correlation functions of the 18-site sample allows for a rough
estimate of the ordered moment of about 60-70\% of the saturation value. Long-wavelength fluctuations
on larger systems might further reduce this value, but with such a large value on 18 sites, we expect
the order to survive in the thermodynamic limit. 
Note that, due to zero modes, estimating the ordered moment
within flavor-wave theory would require pushing the expansion beyond linear order.
So altogether, ED provide very clear evidence in favor of the three-sublattice flavor-wave state.

Let us now briefly discuss the experimental implications of these results. Reaching sufficiently low temperatures is currently a major challenge in ultracold atomic systems
: interestingly,
the exchange integral of the SU(N) case is equal to $2t^2/U$, independently of $N$. It is thus realistic to expect that the exchange scale can be reached for SU(N) fermions as soon as it is reached for SU(2) ones. In that respect, it will be important in experiments to carefully choose the optimal coupling strength $U/t$, which should be large enough
to put the system into the Mott insulating phase described by the SU(3) Heisenberg model, but not too large to
lead to accessible values of the energy scale set by the exchange integral.

Next, we note that the conclusions regarding the selection by quantum or thermal fluctuations for SU(3) fermions are actually valid in any dimension $D\ge 2$, as can be checked easily from Eqs.~(\ref{quantum_modes},\ref{thermal_modes}).
So the present results allow us to make predictions
both for the square and cubic lattices, and the competition between quantum and thermal order-by-disorder should lead
to a rather rich physics. In both 2D and 3D, the system should first develop two-sublattice ordering tendencies
as it is cooled below the exchange scale. In two dimensions, we expect the system to undergo a finite temperature transition at lower temperatures into a directionally ordered state (selection between the two independent $\mathbf{Q}_3^\pm$ spiral propagation vectors), and to reach a three-sublattice ordered state at zero temperature. In three dimensions however, a finite temperature transition into a two-sublattice ordered state is in principle possible, leading
to two possible scenarios: upon lowering the temperature, the system might first undergo a transition into a two-sublattice ordered state, which is followed by a second transition into the three-sublattice ordered state selected by
quantum fluctuations. Alternatively, it could undergo a direct first-order transition from the paramagnetic into the three-sublattice 
ordered state. High-temperature series expansion of the SU(N) case on the 3D cubic lattice seems to favor the second possibility~\cite{fukushima2005}.


Finally, the detection of the three-sublattice order might be attempted using noise correlations~\cite{altman}, 
since the structure factor is expected to have a peak at $\mathbf{Q}_3^\pm$. Alternatively, a recent report of single atom 
resolution experiments~\cite{directImaging} suggests that direct imaging might be possible provided that some contrast can be achieved between different atomic species.

\acknowledgments
 We thank M. Hermele and V. Gurarie for interesting discussions on this subject. This work was supported by the Swiss National Fund, by MaNEP and by the Hungarian OTKA Grant No. K73455. F.M. and K.P. are thankful for the hospitality of the MPIPKS Dresden where this work has been finalized.

\end{document}